\documentclass[12pt]{article}
\begin{document}

\begin{centering}
{\LARGE \bf{Defining civilization utilizing anthropic reasoning}} 
Kevin\ A.\ Pimbblet \\
Address: Department of Physics, University of Queensland, Brisbane, Queensland, QLD 4072, Australia \\
e-mail: pimbblet@physics.uq.edu.au \\
\end{centering}

\section*{Abstract}
We utilize anthropic reasoning to demonstrate that we are typical observers of our 
reference class under a self-sampling assumption by investigating the definition of 
what a civilization is.  With reference to the conflict between such reasoning and the 
observational lack of extra-terrestrial intelligent life, we conclude that a part of our 
theoretical understanding of the Universe will be at fault.

\section{Introduction}

There exist certain physical observations that we should not be at all surprised about.  
What we can expect to observe is restricted a priori by the conditions that are 
necessary for us to exist in the first place$^1$.  Necessarily, if intelligent life forms did 
not evolve in the Universe, then such observations could not take place at all$^{2,3}$.  Yet the 
question of what our own circumstances should be remains a taxing issue$^4$: are we 
typical of all observers in the Universe?  Indeed, Bostrom$^5$ suggests reasoning that we 
should think of ourselves as a random sample derived from the set of all observers in 
our reference class under a `self-sampling assumption'.

One immediate question that arises is how should we define our own reference class?  
Bostrom$^5$ includes all observers who have existed in the past and all those who will 
exist in the future.  Olum$^6$ favours adding to Bostrom's reference class all those 
observers who might potentially have existed as well.  Yet removing these extensions 
and confining our reference class to only those observers who exist presently, there is 
still conflict between anthropic reasoning and observation$^4$ such that we would expect 
ourselves to be part of a titanic (inter-) galactic civilization rather than being part of a 
uniplanetary one.  In numerical terms, only 1 in 100 million individuals would not be 
part of such a civilization at a pan-galactic level$^4$.

In general, the assumptions that underscore this anthropic reasoning are inflationary$^7$: 
in an infinite Universe, there logically exist such titanic civilizations$^{3,4}$.  Given that 
`they' should already be here and are not$^3$, there are a number of scenarios for the 
resolution of this problem$^4$.  In this work, we focus on an observational approach to 
resolving this conflict.

\section{Defining civilization}
Observationally, it is self-evident that we are part of a civilization.  The definition of 
what a civilization is, however, is somewhat tricky and potentially arbitrary.  For 
example, let us consider ourselves as being part of a civilization based on present-day 
national boarders where voluntary migration is uncommon.  Unless we have 
emigrated away from our nation, we would not be surprised to find out that we are 
typical members of said civilization.  Conversely, if we have migrated, we would also 
be unsurprised that we constitute a small minority of the populous in our new 
residential country.

By taking this example further, we can consider our national boundary to be the Earth.  
This in turn expands our reference class to all individuals present on Earth.  Again, we 
would be unsurprised that we are a national of this planet.  Assuming extra-terrestrial 
intelligence does exist, then we may migrate to their planets where we would 
unsurprisingly be in the minority.  Yet we could now consider ourselves to be part of 
some greater civilization with a different (arbitrary) border such as a pan-galactic 
one that consisted of all intelligent life in the Galaxy.  The reference class that we 
now have is exactly as defined in section~1.  We conclude that we would still be 
typical galaxy nationals despite our divergent heritages under the self-sampling 
assumption.  

This line of reasoning indicates a type of selection effect in how civilizations are 
defined.  Specifically, we have used at each step only those observers whom we know 
by direct observation to be existent.  It appears that by only considering our 
observations rather than the (theoretically) implied observations of vast civilizations, 
we are typical observers of our reference class; even when accounting for a finite
speed of light.

\section{Implications}
Returning to our present knowledge base, it is clear that we have only observed our 
own Earthbound civilization.  Further, it is highly probable that there is not any `local' 
extra-terrestrial intelligence waiting to be discovered$^3$.  Importantly, we do not know 
anything about the existence of such beings at larger distances.  The inference that 
they may exist is theoretically based rather than observationally.  There is no 
observational evidence of absence beyond our local stellar region.

We therefore find it probable that it is not anthropic reasoning that is at fault.  
Although more detailed observations of exo-planets will be required, more likely the 
flaw in the conflict lies within a theoretical aspect coupled with the use of theoretical 
observers in our reference class.  Specifically with reference to Olum$^4$, these flaws 
are: few civilizations are able to grow to titanic proportions; the Universe is finite; or 
large-scale inter-stellar colonization is near impossible.

\section{Summary}
We have investigated the definition of civilization and applied anthropic reasoning to 
suggest that we are typical residents of our neighbourhood: whether this is in an 
Earthbound context or spatially more extended.  Thus we reject the supposition that 
anthropic reasoning is invalid and favour that either our theoretical understanding of 
the Universe, our understanding of our colonization ability or the use of theoretical 
observers is at fault.

\section*{Acknowledgements}
KAP thanks the staff at both the University of Durham and the University of 
Queensland for their support. This work was supported by an EPSA University
of Queensland research fellowship.

\section*{References}

(1) B. Carter, Confrontation of cosmological theories with observations (Reidel, Dordrecht), 1974. \\
(2) R.H. Dicke, Nature, 192, 440, 1961.\\
(3) J.D. Barrow and F.J. Tipler, The anthropic cosmological principle, (Clarendon Press, Oxford), 1986.\\
(4) K.D. Olum, arxiv:gr-qc/0303070, 2003.\\
(5) N. Bostrom, Anthropic bias: Observation selection effects, (Routledge, New York), 2002.\\
(6) K.D. Olum, Phil. Q., 52, 164, 2002.\\
(7) A.H. Guth, Phys. Rep., 333, 555, 2000.\\

\end{document}